\documentclass[peerreview]{IEEEtran}
\IEEEoverridecommandlockouts
\usepackage{cite}
\usepackage{amsmath,amssymb,amsfonts}
\usepackage{algorithmic}
\usepackage{graphicx}
\usepackage{textcomp}
\usepackage{multirow}
\usepackage{xcolor}
\def\BibTeX{{\rm B\kern-.05em{\sc i\kern-.025em b}\kern-.08em
    T\kern-.1667em\lower.7ex\hbox{E}\kern-.125emX}}
\begin{document}

\title{Characterizing Deep-Learning I/O Workloads in TensorFlow}

\author{\IEEEauthorblockN{Steven W. D. Chien$^1$, Stefano Markidis$^1$, Chaitanya Prasad Sishtla$^1$, \\ Luis Santos$^2$, Pawel Herman$^1$, Sai Narasimhamurthy$^3$, Erwin Laure$^1$}
\IEEEauthorblockA{
\textit{\\$^1$ KTH Royal Institute of Technology}, Sweden \\
\textit{$^2$ Instituto Superior T\'ecnico}, Portugal\\
\textit{$^3$ Seagate Systems UK}, UK
}
}

\maketitle

\begin{abstract}
The performance of Deep-Learning (DL) computing frameworks rely on the performance of data ingestion and checkpointing. In fact, during the training, a considerable high number of relatively small files are first loaded and pre-processed on CPUs and then moved to accelerator for computation. In addition, checkpointing and restart operations are carried out to allow DL computing frameworks to restart quickly from a checkpoint. Because of this, I/O affects the performance of DL applications. 

In this work, we characterize the I/O performance and scaling of TensorFlow, an open-source programming framework developed by Google and specifically designed for solving DL problems.  To measure TensorFlow I/O performance, we first design a micro-benchmark to measure TensorFlow reads, and then use a TensorFlow mini-application based on AlexNet to measure the performance cost of I/O and checkpointing in TensorFlow. To improve the checkpointing performance, we design and implement a burst buffer.

We find that increasing the number of threads increases TensorFlow bandwidth by a maximum of 2.3$\times$ and 7.8$\times$ on our benchmark environments. The use of the tensorFlow prefetcher results in a complete overlap of computation on accelerator and input pipeline on CPU eliminating the effective cost of I/O on the overall performance. The use of a burst buffer to checkpoint to a fast small capacity storage and copy asynchronously the checkpoints to a slower large capacity storage resulted in a performance improvement of 2.6$\times$ with respect to checkpointing directly to slower storage on our benchmark environment.
\end{abstract}


\section{Introduction}

Deep-learning (DL) frameworks are increasingly taking advantage of HPC systems with accelerators to speed-up training of large and complex neural networks. During a training process, connection weights in a network are iteratively adjusted to minimize a loss function which reflects the error of a network. This process involves feeding the network with large amount of training samples which implies computationally intensive matrix multiplications. For this reason, accelerators such as GPUs are commonly used. Similar to many HPC applications, I/O is one of the major performance bottlenecks as training process requires ingestion of large amount of training samples. Training datasets typically contain several Gigabytes and up to several Terabytes of data. For instance, the YouTube-8M Kaggle challenge provides 1.71 TB of frame-level data\footnote{https://www.kaggle.com/c/youtube8m}. However, the size of individual training samples is usually limited. The Large Scale Visual Recognition Challenge 2013 (LSVRC2013)~\cite{ILSVRC15} training dataset contains 395,909 images, yet the average image resolution is only 482$\times$415 pixels\footnote{http://www.image-net.org/challenges/LSVRC/2013/}. With ever faster accelerators for computation~\cite{jouppi2017datacenter}\cite{markidis2018nvidia}, it is important to maintain a high data ingestion rate such that devices will not be kept idle due to slow I/O.

TensorFlow is among the most successful open-source frameworks for neural networks development~\cite{abadi2016tensorflow}. The TensorFlow GitHub repository\footnote{https://github.com/tensorflow/tensorflow} shows more than 1,600 active contributors with an average of more than a thousand commits per month, making it one of the fastest growing computing framework. TensorFlow is accelerator-centric: neural networks are expressed as computation graphs where nodes represent operations and edges represent data consumed or produced by operations. These operations can be easily offloaded to different specialized accelerators for DL workload~\cite{jouppi2017datacenter}\cite{markidis2018nvidia} and computation will be completely performed on these devices. CPUs on the host system, on the other hand handle all I/O operations such as loading of training or validation samples from files to memory and checkpointing for restart.

TensorFlow attempts to alleviate the I/O bottleneck by performing independent reading of these training samples in parallel through an input pipeline. This reduces idling time of accelerators due to I/O through better utilization of bandwidth thus increases ingestion rate. For large systems with a large number of computing nodes, failure is becoming increasingly common and failure recovery must be properly handled~\cite{node-failure}. DL training is no exception. During a long running training process, snapshots of weight variables can be taken periodically to allow later restarting of training in case of system failure. TensorFlow implements this as a checkpoint feature and saves the state of variables of a computation graph to disk. The file size of these checkpoints depend on the network in use and each checkpoint can easily reach hundreds of Megabytes size in the case of most modern architectures.

For all these reasons, it is important to characterize the impact of I/O on training performance and optimize accordingly. In particular, we study performance of the I/O interface provided by TensorFlow: Dataset API. Furthermore, we study how emerging storage technologies such as non-volatile memory can help eliminate bottleneck due to slow I/O. We first develop a benchmark tool to measure raw performance of the data I/O pipeline. We proceed to develop a mini-application for image classification with AlexNet, which illustrates typical computation of a simple neural network with a data ingestion pipeline. Finally, we extend the mini application with checkpointing capability and implement a proof-of-concept burst buffer for temporary staging of checkpoints. With these tools, we evaluate experimentally the I/O performance of TensorFlow.

The contributions of this work are:
\begin{itemize}
\item We design and implement a TensorFlow I/O benchmark, mimicking the STREAM benchmark, to evaluate the bandwidth of TensorFlow I/O on HPC systems.
\item We design and implement a TensorFlow mini-application to mimic the typical input data pipeline and computation in a DL application. We use an AlexNet network with a batches of images as input and estimate the impact of the I/O pipeline.
\item We evaluate the impact of multi-threading, prefetching and caching on the performance of the micro-benchmark and mini-application.
\item We extend the mini-application with checkpointing capability and implement a proof-of-concept burst buffer where checkpoints are staged in non-volatile memory and evaluate checkpointing performance comparing to difference storage devices.
\end{itemize}

The paper is organized as follows. Section~\ref{overview} provides an overview of the I/O system in TensorFlow. We describe the design and implementation of the benchmark and mini-application in Section~\ref{methods}. Results are presented in Section~\ref{results}. Finally, Section~\ref{discussion} concludes the paper, summarizes and discusses the results. 
\section{Overview of TensorFlow I/O}\label{overview}
TensorFlow is an open-source programming framework developed by Google specifically designed for solving deep-learning problems. Application developers can implement neural networks on TensorFlow with a variety of APIs of different language bindings, such as Java, Go and Rust. The most complete API is the Python API and is used by the majority of TensorFlow users. For this reason, Python is the language of choice for the tools developed in this paper.

\begin{figure}[t]
	\begin{center}
		\includegraphics[width=0.7\columnwidth]{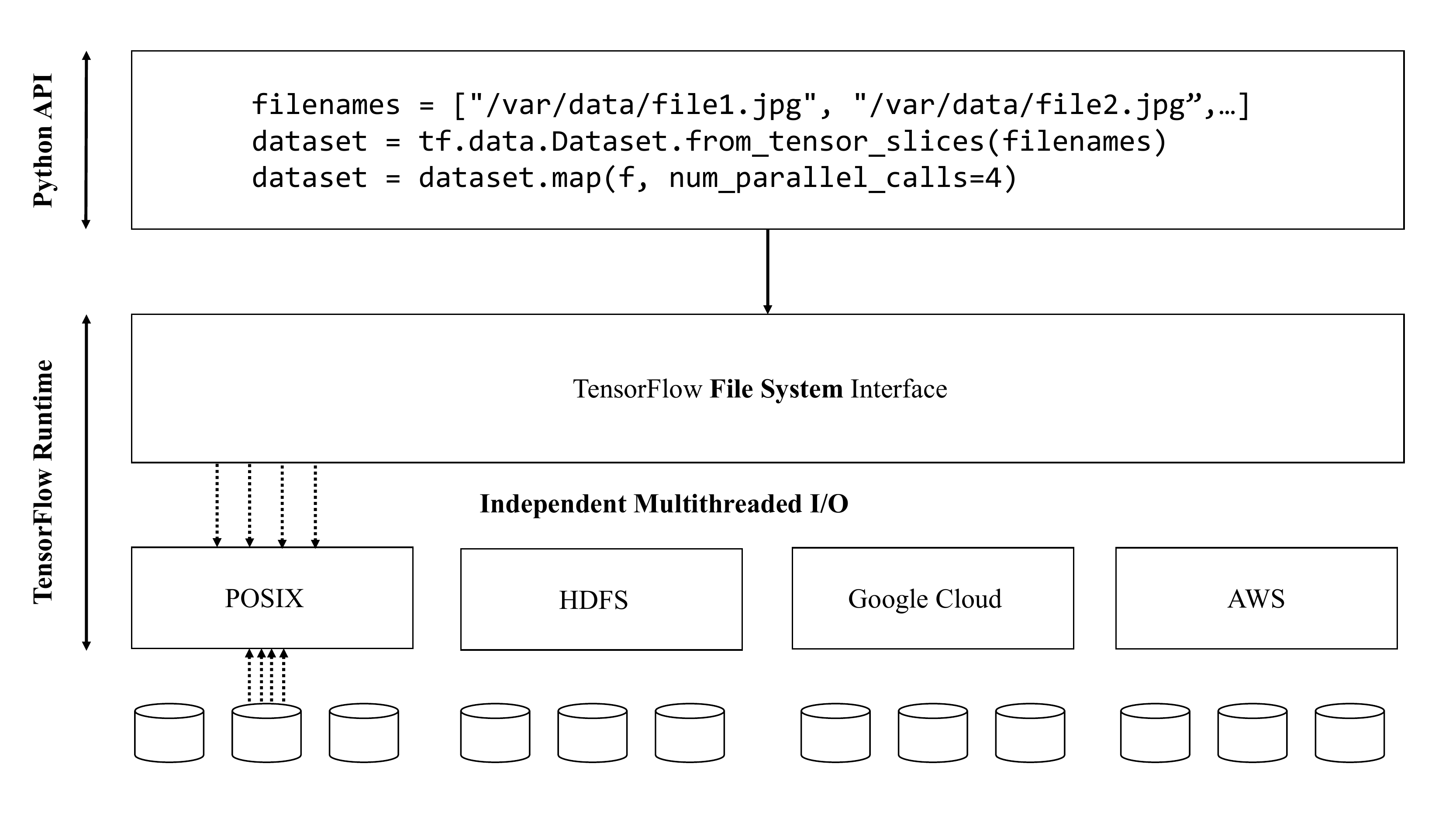}
		\caption{TensorFlow is able to interact with different types of storage systems, including POSIX file system, Hadoop Distributed File System, Google Cloud Storage and Amazon S3 through file system adapters. Through parallel mapping or parallel interleaving it is possible to read files concurrently.}
		\label{filesystems}%
	\end{center}
\end{figure}

Operations in TensorFlow are powered by the TensorFlow runtime which is written in C++. The runtime provides mechanisms for managing and scheduling data movement across hosts and devices, offloading computation to accelerators, memory allocation and most importantly for this paper: I/O operations. Fig.~\ref{filesystems} illustrates the design of TensorFlow file system module. Currently, TensorFlow 1.10 supports four underlying file systems, namely: traditional POSIX file system; S3, a cloud based object storage system by Amazon Web Service; Google Cloud Storage, another cloud based file system by Google; and Hadoop Distributed File System (HDFS). Different I/O adapters are developed such that users can interact with these file systems through the TensorFlow API. These adapters inherit the same interface for easy switching and extension from POSIX I/O to other I/O systems. For example, the POSIX adapter implements writing of file through \textsf{\small{fwrite()}} and reading through \textsf{\small{pread()}}. From a user's point of view, the switching of a file system can be easily done by substituting the prefix of a file path.

TensorFlow also provides distributed runtime support which enables applications to run on clusters and HPC systems. Communication in distributed TensorFlow is based on gRPC, an open-source RPC (Remote Procedure Call) library developed by Google. However modules are developed to support different transport layers such as InfiniBand RDMA (Remote Direct Memory Access) and MPI.

\subsection{Input Pipeline}
TensorFlow neural network training follows a data-driven programming model which is similar to that of a producer-consumer model. As soon as the accelerators complete a training step (consumer), it ingests training samples from the input pipeline for computation (producer). The input pipeline typically performs actual I/O, decoding and pre-processing. One example would be the ingestion of JPEG images where the images are read from files. These image data are decoded into tensors and pre-processing e.g. resizing, is applied to suit the input size of a given neural network.

The TensorFlow Dataset implementation attempts to optimize this process by overlapping the input pipeline with the computation pipeline. A common approach in neural network training is the use of mini-batches. Since a training set typically does not fit into GPU memory, training is instead done in batches per iteration. When the computation pipeline becomes ready for the next iteration, one new batch is ingested from the input pipeline to the computation pipeline. By overlapping the preparation of batches with the computation, new batches can be buffered and become ready for ingestion as soon as the computation pipeline becomes available. More specifically, the input pipeline can proceed to prepare the next batch as soon as the current batch is completed, rather than waiting for the computation pipeline to request for a new one. Furthermore, since the input pipeline operates on CPU and computation is typically done on GPU, this allows efficient utilization of both devices on a machine.

\begin{figure}[t]
\begin{center}
\includegraphics[width=0.6\columnwidth]{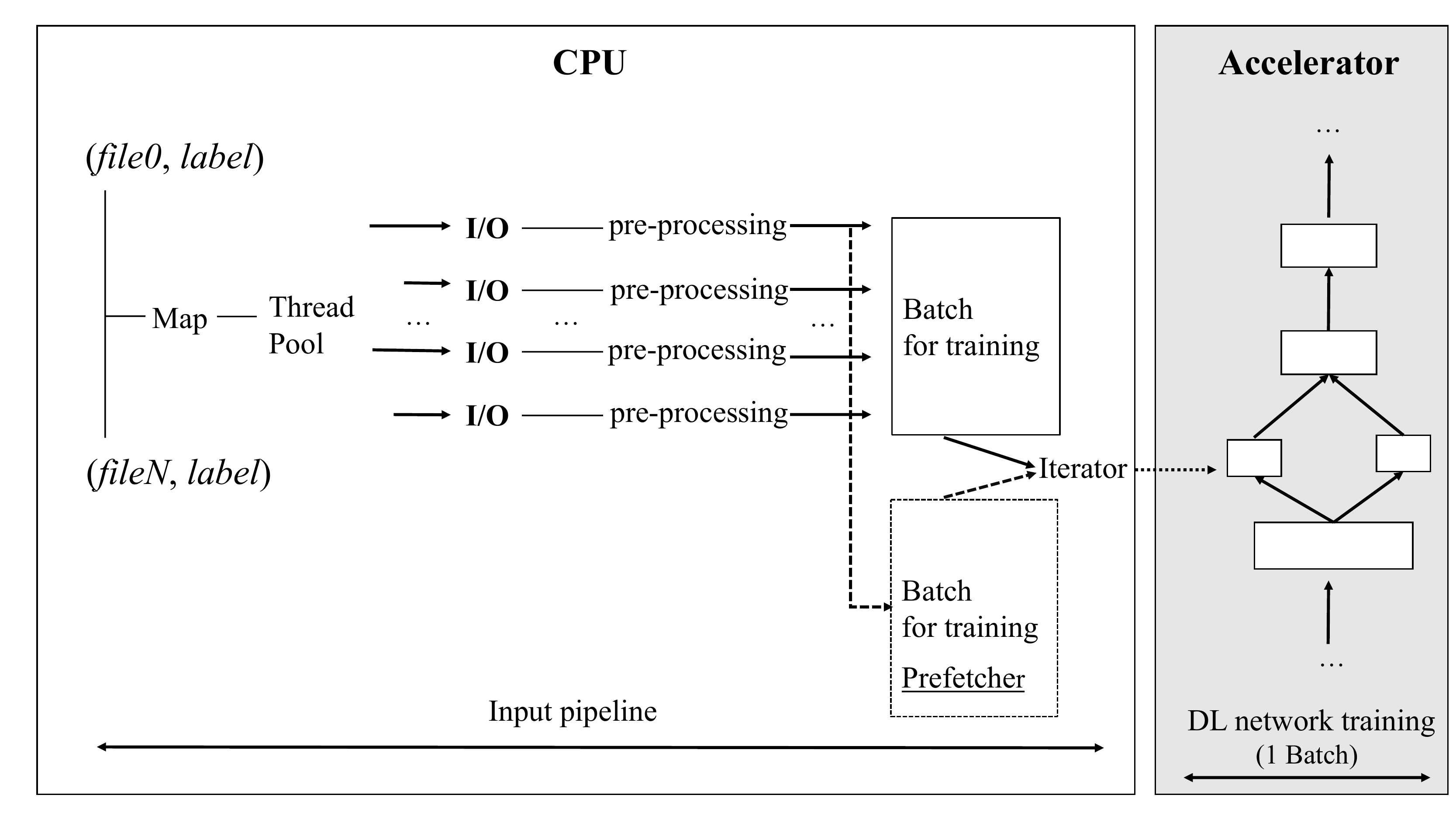}
\caption{A simplified TensorFlow input pipeline. In a typical use case, a list of file paths and their labels form a source dataset. The file names are sent to a transformation function through map to perform actual I/O and decoding. Finally these transformed data from mapping are collected into a batch. Optionally a prefetch operation can be used to cache one or more batches in memory such that they can be sent through the iterator immediately when requested.}
\label{iopipeline}%
\end{center}
\end{figure}

\subsubsection{Parallel I/O}
The reading and pre-processing is applied to individual files with individual outputs. In other words, this is an embarrassingly parallel process. Therefore, it is possible to parallelize the I/O of many files by mapping a list of file names for transformation in form of \textsf{\small{tf.dataset.map()}}. Threads will be spawned by the runtime to execute the I/O function and the number of threads used by the map can be specified with \textsf{\small{num\_parallel\_calls}} argument. The result from mapping will be collected by the down stream operation. The user can optionally apply shuffling of the dataset by \textsf{\small{tf.dataset.shuffle()}} and specify the size of the shuffle buffer. The \textsf{\small{tf.dataset.batch()}} can be applied by specifying the batch size. This operation accumulates the number of training samples from upstream operation and form a batch.

\subsubsection{Prefetching}
In order to overlap the computation pipeline and input pipeline, \textsf{\small{tf.dataset.prefetch()}} can be used to specify how many batches of data to be buffered. This ensures that there will be a specified number of batches ready for consumption. \textsf{\small{tf.data.Dataset.prefetch()}} transformation is used to prefetch data from the previous pipeline operation. When applied after \textsf{\small{tf.dataset.batch()}}, this allows buffering of one or more batches which are ready for direct consumption. The TensorFlow runtime implements a prefetcher as a background thread and a consumption function. The thread maintains a buffer which stores elements that are prefetched from the upstream operation. The buffer uses a double ended queue implementation from standard library. The thread itself contains an infinite loop which waits for a condition variable. When a Tensor is consumed from the buffer using a consumer function, the thread is notified through the condition variable and wakes up to fetch another element from upstream.

A simplified input and training pipeline is presented in Fig.~\ref{iopipeline}. In the \textsf{\small{tf.data}} API, \textsf{\small{tf.data.Dataset}} represents a container of Tensor objects. The input pipeline itself is constructed with a list of operations provided by the \textsf{\small{tf.data}} API. For instance, a source dataset can be a vector of file name of samples and their respective labels for a classification problem. The dataset can fetch one slice of the vector into the pipeline with \textsf{\small{Dataset.from\_tensor\_slices()}}. These samples can be read and decoded for the next stage of the pipeline by applying a transformation function. This function obtains the slice of information given and returns another slice of transformed data to the downstream operation. One example is to read a file through the TensorFlow file system API, decode data and apply pre-processing. When given the file name of a JPEG image, the transformation function reads it with \textsf{\small{tf.read\_file()}} and decodes with \textsf{\small{tf.image.decode\_jpeg()}}, then converts the decoded image into a tensor with \textsf{\small{tf.image.convert\_image\_dtype()}} and finally applies resizing with \textsf{\small{tf.image.resize\_images()}}. The function returns the transformed image in form of tensor to the next operation, which accumulates to a batch of data. One or more batches are prefetched and buffered by a prefetcher and finally consumed by an iterator.

\subsection{Checkpointing and Restart}
TensorFlow uses the \textsf{\small{tf.train.Saver()}} to checkpoint training sessions to disk and allows restarting computation from a given checkpoint. If no argument is provided to \textsf{\small{tf.train.Saver()}}, all variables in the computation graph are checkpointed. Each tensor is saved under the name that was given when the tensor was created. When a checkpoint is being saved, the checkpoint saver generates three files: a metadata file with suffix .meta which stores the structure of the graph, an index file with suffix .index which describes the tensors of a graph and a data file with suffix .data which stores actual data in variables. When a checkpoint is being restored, the structure of the graph is first restored followed by the restoration of variable contents into the graph. \textsf{\small{tf.train.Saver()}} is responsible both for checkpointing and restoring the training state. The Saver can additionally perform cleanup of old checkpoints. By default only the first five checkpoints are retained.
\section{Methodology}\label{methods}
In this section, we present the design and development of benchmark and mini-applications to benchmark I/O performance of TensorFlow.

\subsection{TensorFlow I/O Micro-Benchmark}
We develop a STREAM-like benchmark application~\cite{stream} to measure the performance of I/O with TensorFlow dataset API. We achieve this by loading a set of images from storage, and perform simple pre-processing. When the application is initialized, a number of user configurable I/O parameters are evaluated. These include batch size and the number of current I/O threads to use. A file containing a list of paths to images to be consumed is also provided which will be loaded as a Python list. This list becomes the source element.

We first apply shuffle transformation to the list to randomize the order of elements, then immediately apply map transformation, where slices of the image path list will be given to a transformation function. The number of threads to perform concurrent mapping is set according to the parameter provided by the user. This function performs I/O with \textsf{\small{tf.read()}} and data from read is immediately passed to \textsf{\small{tf.image.decode\_jpeg()}} followed by a resize operation \textsf{\small{tf.image.resize\_images()}}. A \textsf{\small{tf.contrib.data.ignore\_errors()}} is applied after mapping to avoid exceptions in the mapped function from terminating all execution. This is useful in processing large amount of data where data completeness is not guaranteed. The tensors returned by this transformation through mapping will be collected by a batch transformation where tensors will be accumulated until a batch is formed. The batch size is set according to user configuration. Finally an iterator is attached to the end of the dataset. We directly invoke the iterator of the dataset in a session to draw new batches without passing them through any computation phase. The iterator is continuously invoked until a defined number of batches is consumed. By measuring the time between the first invocation and last invocation it is possible to compute the ingestion rate with the number of images passing through the iterator and the number of seconds used. We also perform another execution with a extremely simplified input pipeline where all steps are omitted in the mapped function except \textsf{\small{tf.read()}} to isolate effect due to preprocessing.

\subsection{TensorFlow Mini-Application: AlexNet}
We develop a mini-application using the AlexNet DL network architecture to mimic the typical DL I/O workload. We choose AlexNet because it is easy to implement (roughly 200 lines of code) yet it retains basic characteristics of DL networks. AlexNet was developed by Alex Krizhevsky (hence the name), and was winner of the ImageNet challenge in 2012. AlexNet~\cite{krizhevsky2012imagenet} consists of five convolution layers, three max pooling layers and three fully connected layers. The network uses the ReLU (Rectified Linear Unit) activation function and the convolution layers contribute to feature extraction while the fully connected layers implement the classifiers.

\begin{figure}[t]
	\begin{center}
		\includegraphics[width=\columnwidth]{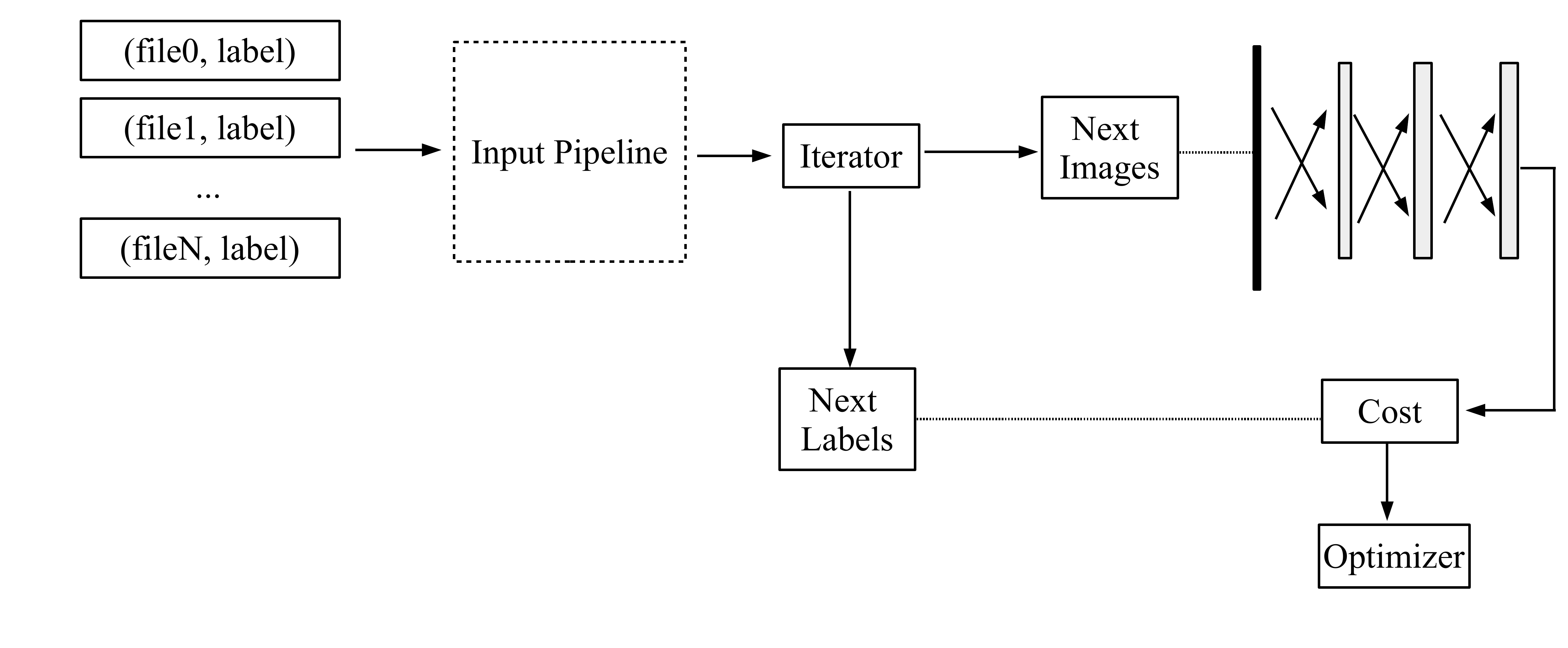}
		\caption{Workflow of AlexNet mini-application. Image files are ingested to form batches of image data and labels through an input pipeline. The neural network takes a batch of image data and perform classification through AlexNet. The output of the classifier is compared with the ground truth (labels) and a cost value which represents the error of classification is derived. The cost value is passed to a TensorFlow optimizer, which base on a minimization target adjusts variables in the network with DL methods. The mini-application executes one training step and consume one batch of images by calling the optimizer.}
		\label{miniapp-dataflow}%
	\end{center}
\end{figure}

During initialization, the user provides a list of paths to images and labels, batch size, total number of batches per epoch, number of I/O threads to use and indicates if prefetch is to be enabled or not. The input pipeline consists of three stages. Similar to the micro-benchmark application, we read the list of image paths and their classes. The labels are converted to one-hot vectors to align the number of outputs of the classifier. We apply shuffle transformation to the list of samples to randomize order of elements. We also perform map transformation where paths will be sent to a transformation function. Again, the number of threads used to perform concurrent mapping is set according to the parameter provided by the user. This function performs I/O with \textsf{\small{tf.read()}} and data from read is immediately passed to \textsf{\small{tf.image.decode\_png()}} followed by a resize operation \textsf{\small{tf.image.resize\_images()}}. \textsf{\small{tf.image.decode\_png()}} supports both decoding of JPEG and PNG images. In our implementation, input images are resized to dimensions 224$\times$224 with three channels. The tensors returned by this transformation through mapping will be collected by a batch transformation where tensors will be accumulated until a batch is formed. The batch size is set according to user configuration. Finally an iterator is attached to the end of the dataset.  We derive a cost value from the network to reflect the error from classification and pass it to the Adam Optimizer provided by TensorFlow. In each invocation, a batch is drawn from the dataset through the neural network to compute the cost value and the optimizer minimizes the cost by modifying weights variables of the network accordingly. A simplified work flow of the mini-application is presented in Fig.~\ref{miniapp-dataflow}.

\subsection{Checkpoint and Burst Buffer}
We extend the AlexNet mini-application to add checkpointing capability with \textsf{\small{tf.train.Saver()}}. A checkpoint prefix and the interval between checkpoints are defined by the user. A checkpoint prefix refers to the path and prefix name used for the checkpoint file. For every defined number of iterations, the checkpoint saver will be called to take a snapshot of all variables. In order to ensure that the checkpoints are safely written to disk, we perform disk synchronization of disk with \textsf{\small{syncfs()}} immediately after Saver returns.

In order to take a snapshot of all variables, training process must be paused in favor of variable saving process. A simple network such as AlexNet in our mini-application can produce up to several hundreds of megabytes of snapshot files. The process of writing these snapshot files seriously hamper training performance. For this reason, we introduce a simple proof-of-concept burst buffer implementation in our extended mini-application. In particular, we take advantage of non-volatile memory storage device for fast writing of snapshot files. 

Non-volatile memory is an emerging storage technology which features access speed while being persistent~\cite{peng2016exploring}. For this reason, it has been increasingly adopted as a burst buffer, which acts as an intermediary between slower but larger; and faster but smaller storage devices to absorb bursty I/O which affects application performance. Burst buffers are increasingly adopted in large scale cluster systems~\cite{bhimji2016accelerating}. Two of the most famous burst buffer solutions are DataWarp~\cite{henseler2016architecture} by Cray and IME~\cite{schenck2017evaluation} by DNN. In DataWarp, user requests resources from the system through job scripts. After allocation the system can be interacted through POSIX APIs. Data movement can be requested explicitly when used as a scratch drive or done implicitly when used as a cached instance. IME is designed as an intermediate storage layer between a compute system and an external storage system, connected through switches. The storage system and IME uses NVM-based devices to absorb I/O traffic between the external storage system and compute system.

Checkpointing is a suitable use case for using burst buffer as staging~\cite{sato2012design}. Since I/O due to checkpointing is bursty, by absorbing the writing with a fast storage it is possible to minimize disruption to execution. In our implementation, when the checkpoint saver is called, a checkpoint is created and synchronized to a fast non-volatile memory device. At the same time a process is spawned in background to copy the just created files to hard disk for storage. Since the checkpoint was already written to persistent memory, it is possible to continue training without disruption.
\section{Experimental Environment}\label{experiments}
The first benchmark environment is a workstation with local storage two Hard Disk Drives (HDD) and one Solid-State Drive (SSD) and one Intel Optane card (Opt.). The second benchmark environment is a supercomputer at KTH Royal Institute of Technology, with network-based storage provided by a Lustre parallel file system. 

The two systems have the following characteristics:
\begin{itemize}
\item{\bf Blackdog} is a workstation with an eight core Xeon E5- 2609v2 processor running at 2.5GHz and an NVIDIA Quadro K4000. The workstation is equipped with a total of 72GB DRAM. The storage consists of two 4TB HDD (non-RAID) and a 250GB SSD (Samsung 850 EVO). The machine also has one Intel Optane SSD 900p with 480GB on PCIe. The file system used is ext4 and OS is Ubuntu Server 16.04 with Kernel 4.4.0-112-generic. TensorFlow version 1.10 was compiled with gcc v7.3.0 and NVIDIA CUDA 9.2. 

\item {\bf Tegner} is a GPU cluster at KTH PDC. Each node has two Intel E5-2690v3 Haswell processors, 512 GB of RAM and one Nvidia Quadro K420 or K80 GPU and provides 1GB and 24GB of ram respectively. The parallel file system used is Lustre and operating system is CentOS 7.4.1708. The nodes are connected by EDR InfiniBand network. We compiled TensorFlow 1.10 with support of Python 3.6, NVIDIA CUDA, OpenMPI and InfiniBand. The versions used are CUDA 9.1, cuDNN 7.0.5, OpenMPI 3.0 and the compiler used is GCC 6.2.0.
\end{itemize}

\begin{table}[ht]
	\centering
	\caption{IOR benchmark results}
	\begin{tabular}{|l|l|l|l|}
		\hline
		\textbf{Platform} & \textbf{Devices} & \textbf{Max Read} & \textbf{Max Write}\\
		\hline
		\multirow{3}{*}{\textbf{Blackdog}}& & &\\
		& HDD & 163.00 MB/sec & 133.14 MB/sec \\
		& SSD & 280.55 MB/sec & 195.05 MB/sec \\
		& Intel Optane & 1603.06 MB/sec & 511.78 MB/sec \\
		\hline
		\multirow{3}{*}{\textbf{Tegner}}& & &\\
		& Lustre & 1968.618 MB/sec & 991.914 MB/sec \\
		\hline
	\end{tabular}
	\label{ior-results}
\end{table}

Benchmarking I/O is a notoriously difficult task. The reason is that the Operating System often caches files which are previously opened in memory, thus the real cost of I/O cannot be assessed. For this reason, we used a small C program which passes a POSIX advice with \textsf{\small{posix\_advice()}} for the files used with \textsf{\small{POSIX\_FADV\_DONTNEED}}, which states that the file is no longer needed. Furthermore, on our local workstation where we have roots privileges we use \textsf{\small{\# echo 1 $>$ /proc/sys/vm/drop\_caches}} to instruct the operating system to drop caches. We also monitor I/O activities using \textsf{\small{iosnoop}} and \textsf{\small{dstat}} tools to ensure I/O activities actually occurred. On the clusters, we only execute the small C program to pass POSIX advice as access to root privilege is not possible. In order to establish a upper bound of I/O performance on our test platform, we summarize IOR benchmarking results in Table~\ref{ior-results}. We obtain this result by reading from and writing to a file with 5GB on the respective devices for six times and obtain the bandwidth. The execution run is for warm up and the result is discarded. The median bandwidth is reported here. Again caches are dropped before the tests.

\subsection{TensorFlow I/O Micro-Benchmark}
We execute our micro-benchmark tool with a subset of images from ImageNet totaling 16,384 JPEG images with median image size 112KB. We choose the particular batch size 64 to present the test results. Therefore, we invoke the iterator through a TensorFlow session 256 times to fully consume the images. We perform strong scaling by varying the number of threads used for map transformation, which are responsible for performing individual I/O of images. The number of threads used are one, two, four and eight and we report the bandwidth with number of images processed per second. On our local workstation we repeat the tests with sample images placed on different devices: HDD, SSD and Intel Optane. On Tegner, we perform the tests with sample images placed on the parallel file system Lustre. We execute each test six times, with the first test being warm up and report the median bandwidth value. Error bars are not shown as bandwidth measurements vary less than 1\% and 4\% with the respect to median bandwidth on the Blackdog and Tegner respectively.

\subsection{TensorFlow Mini-Application: AlexNet}
We perform the tests with Caltech 101 data set, a small scale data set with 9,144 images of 101 classes plus one extra \emph{Google background} class. The median image size is approximately 12kB while the average size is around 14kB. We report our results based on batch size 64 where our GPU is well utilized. Since 9,144 is not divisible by 64, in each test we execute 142 iterations which in total process 9,088 images. In order words, we only execute one epoch of training. The reason is that after the first epoch all samples will be seen by the Operating System and will potentially be cached in memory, thus avoiding actual I/O operations. We vary the number of threads used for mapping between one, two, four and eight and report the runtime in seconds. We additionally test for the effect with and without prefetching by setting the number of batches to prefetch to one or zero batch. Similar to the tests with micro-benchmark tool, we repeat the tests when sample images are placed on different storage devices: HDD, SSD and Intel Optane on our local workstation. On Tegner, the images are placed on Lustre. We execute each test six times, with the first test being warm up and report the median run time.  Error bars are not shown as execution time measurements vary less than 1\% and 6\% with the respect to median bandwidth on the Blackdog and Tegner respectively.

In order to understand the effect of prefetching, we use \emph{dstat}\footnote{http://dag.wiee.rs/home-made/dstat/} to perform tracing of system I/O activities. \emph{dstat} is a system resources monitoring tool which is able to produce statistics on different system activities. Statistics are sampled once per second and can be reported as a comma separated values file (CSV). In particular, we trace for disk activities on devices where samples are placed. We report disk activities from the first and the last invocation of the iterator.

\subsection{Checkpoint and Burst Buffer}
We execute the same tests as used with our mini-application, except that we only perform the first 100 iterations instead of full 142 iterations. We perform a checkpoint every 20 iteration and call for file system synchronization to ensure the checkpoint files are safely written to disk. In this particular tests, we use batch size 64 where images are placed on SSD and prefetch is enabled. We test for checkpoint to HDD, SSD, Intel Optane, and Intel Optane as burst buffer. The total runtime and median checkpoint in seconds are reported in addition to the disk activities measured using \emph{dstat}.
\begin{figure}[t]
	\begin{center}
		\includegraphics[width=0.5\columnwidth]{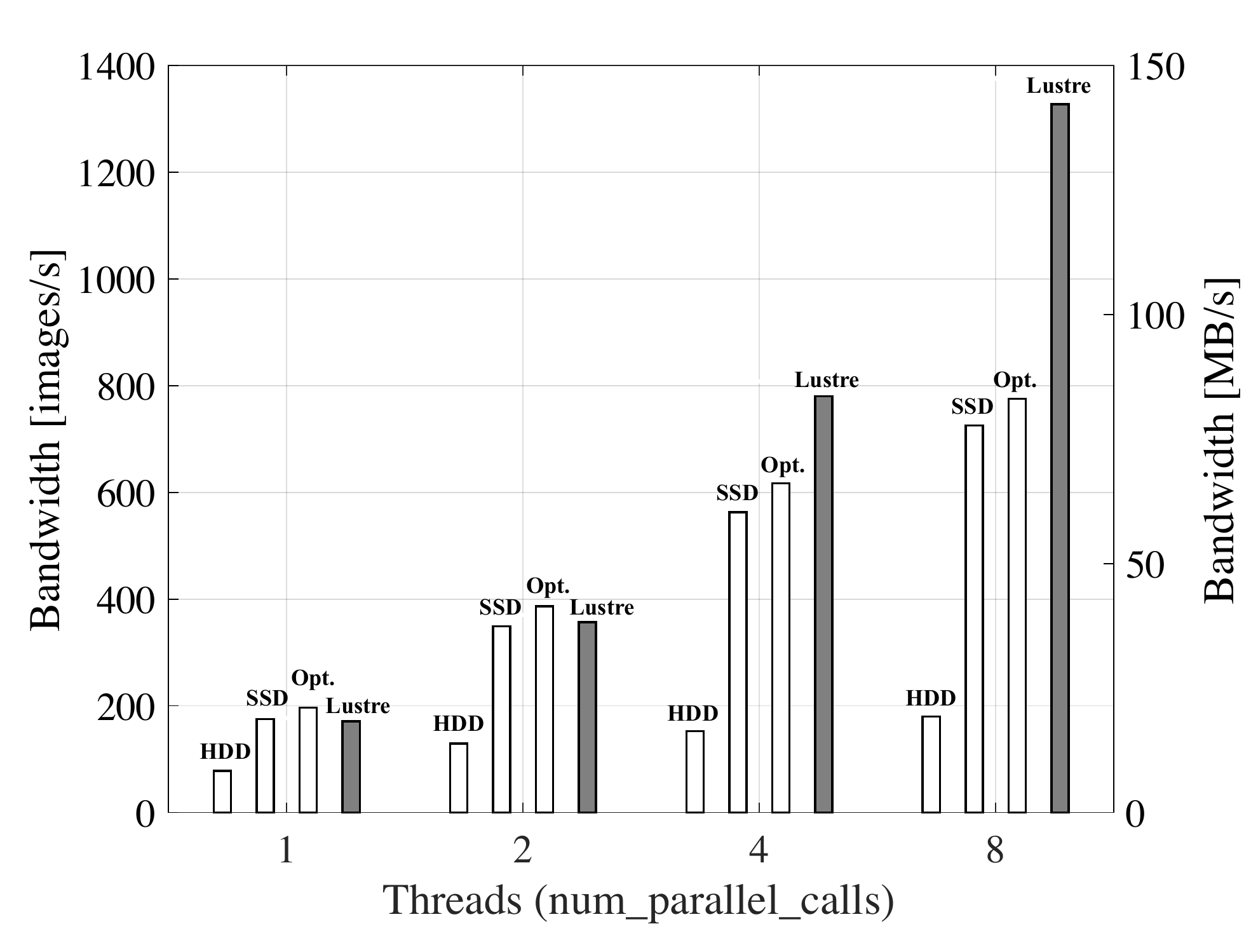}
		\caption{Median bandwidth in images per second from strong scaling tests with micro-benchmark application using a typical input processing pipeline for images. The higher the rate the better. The bandwidth is translated to MB/s scale on the right. Tests that are performed on Blackdog are colored in white and the test conducted on Tegner is performed in gray.}
		\label{micro-benchmark-with-preprocess}%
	\end{center}
\end{figure}

\section{Results}\label{results}
In this section, we present the results of parametric study described in Section~\ref{experiments}. The tests are performed on Blackdog and Tegner.

\subsection{TensorFlow I/O Micro-Benchmark}

\begin{figure}[t]
	\begin{center}
		\includegraphics[width=0.5\columnwidth]{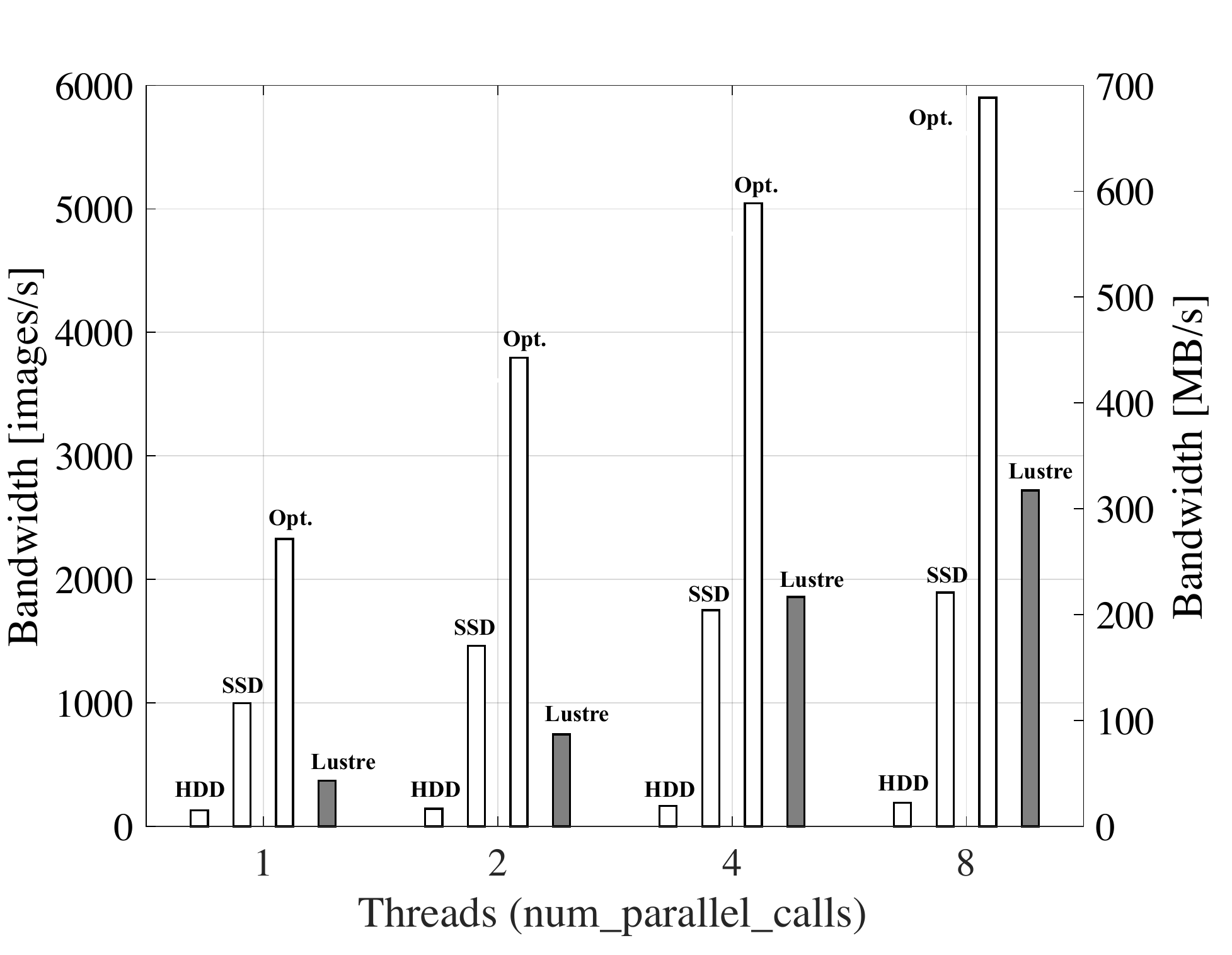}
		\caption{Median bandwidth in images per second when the input pipeline of the micro-benchmark application contains simply a read file operation without further preprocessing. The bandwidth is translated to MB/s scale on the right. Tests that are performed on Blackdog are colored in white and the test conducted on Tegner is performed in gray.}
		\label{micro-benchmark-without-preprocess}%
	\end{center}
\end{figure}

Threading is an effective way of increasing bandwidth utilization for file ingestion. Fig.~\ref{micro-benchmark-with-preprocess} shows that by increasing the number of threads from one to two it effectively almost double the bandwidth on most devices. The effect is particularly visible on fast storage devices such as SSD and Optane. On HDD the effect of scaling is also visible but soon flattens when the number of threads used exceeds four. A not very surprising result is that scaling on Tegner with Lustre file system shows the best scalability when the number of threads increases as the files are likely shared on different object store targets, thus allowing better bandwidth utilization. The bandwidth when comparing to I/O performance measured with IOR is unfavorable. One reason is due to preprocessing functions such as decoding used inside the mapped function which uses computation. Fig.~\ref{micro-benchmark-without-preprocess} shows the performance of an input pipeline without any preprocessing, where the only operation is to read the file with the given path.

\subsection{TensorFlow Mini-Application: AlexNet}

\begin{figure}[t]
	\begin{center}
		\includegraphics[width=0.5\columnwidth]{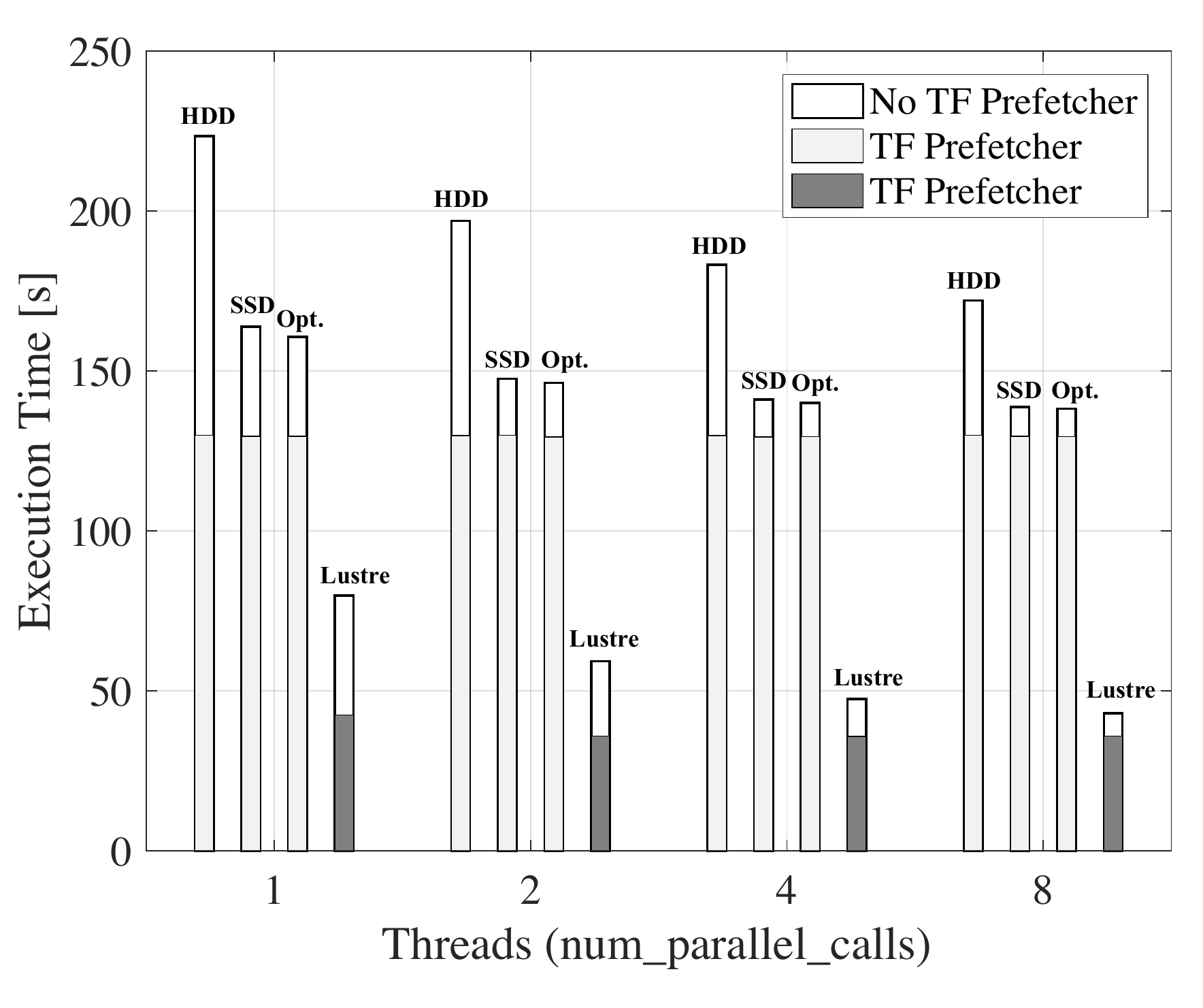}
		\caption{Median execution time of mini-application in seconds with different number of threads when prefetch is disabled or enabled. Tested storage devices where images are stored are HDD, SSD and Optane. Gray bar represents runtime when one batch is prefetched and the white area represents excess in runtime when prefetch is not used. The case of prefetching of one batch on Tegner is shown in dark gray.}
		\label{miniapp-benchmark}%
	\end{center}
\end{figure}

\begin{figure}[t]
	\begin{center}
		\includegraphics[width=0.5\columnwidth]{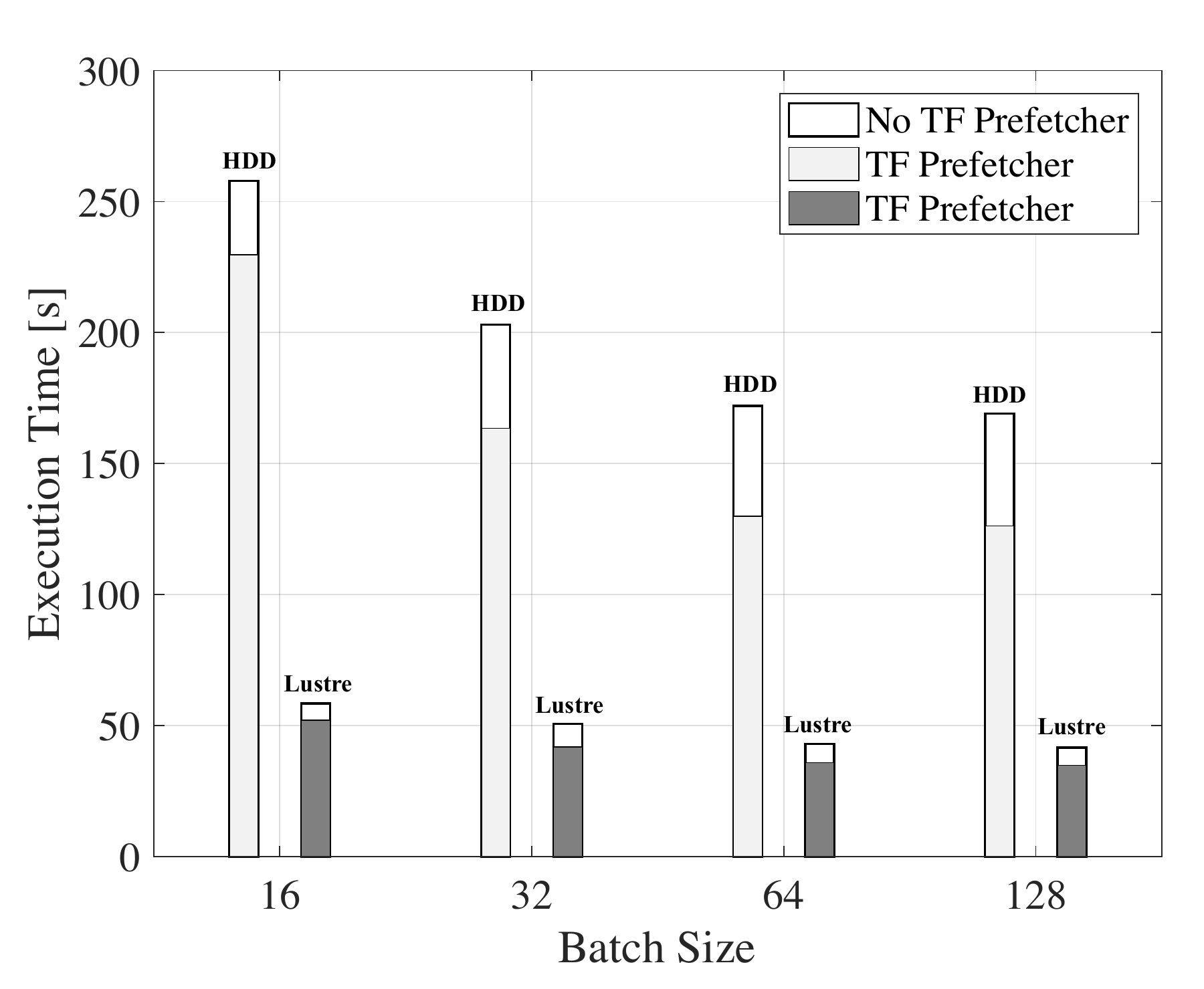}
		\caption{The variation of batch size can impact training performance. A larger batch size tends to allow better utilization of GPUs. The figure shows median training performance of mini-application with eight parallel mapping threads.}
		\label{miniapp-batch}%
	\end{center}
\end{figure}

The impact of overlapping input pipeline on CPU with computation pipeline on GPU is extremely effective. We show that runtime of the mini-application with prefetch enabled in light gray in Fig.~\ref{miniapp-benchmark} while the white portion of the bars represents the excess in runtime when prefetch is not enabled. It can be seen that scaling of I/O thread has moderate effect on the total runtime. This is particular visible on HDD. An interesting observation is that the delay due to I/O can be completely hidden by prefetching. This is indicated by the execution time for different configurations which becomes the same regardless of the number of threads or storage technology used. In other words, the excess in execution time when prefetch is disabled can be conceived as the cost of I/O.

\begin{figure}[t]
	\begin{center}
		\includegraphics[width=0.5\columnwidth]{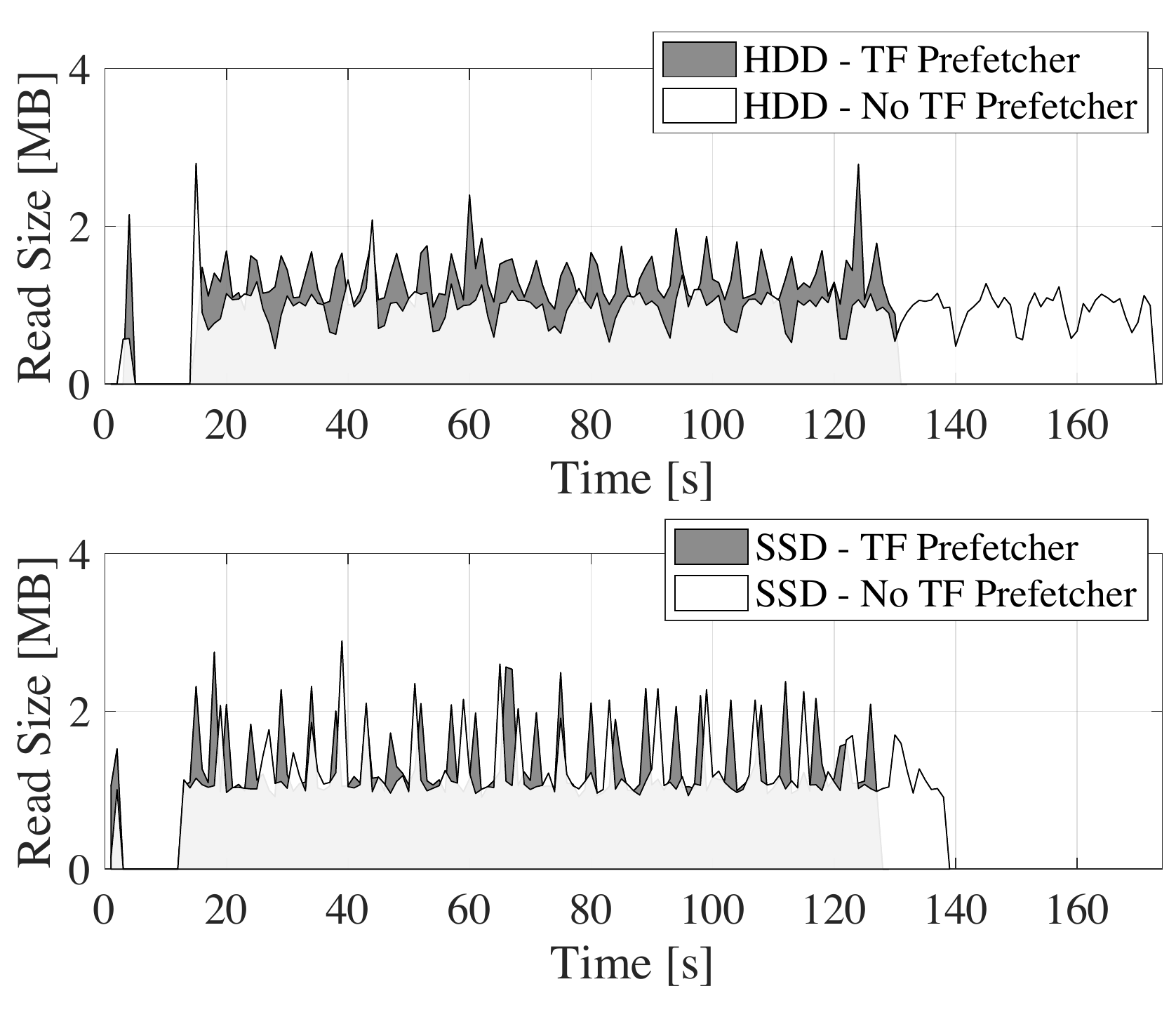}
		\caption{I/O tracing of mini-application in Megabyte over time using \emph{dstat}. Area in gray represents amount of data read in MB read when one batch is prefetched and area in white represents data read when prefetch is not used. Panel on top presents result when samples are placed on HDD and panel on the bottom shows result when samples are placed on SSD.}
		\label{miniapp-tracing}%
	\end{center}
\end{figure}

In order to confirm this observation, we trace I/O activity from the execution of the first to the last with \emph{dstat} in Fig.~\ref{miniapp-tracing}. The white area represents the number of megabytes read each second when prefetch is disabled while the gray area represent results when one batch is prefetched. The panel on the top shows the tracing of execution when samples are placed on HDD while the lower panel shows the case when samples are placed on SSD.

For the case when image files are read from HDD without prefetch, a stable interleaving pattern can be observed between batching drawings. As for the case when one batch is prefetched, intervals are considerably closer and the number of megabyte being read is higher in general.

For the case when images are being read from SSD without prefetch, a similar pattern can also be observed. When one batch is prefetched, high peaks can be observed. However, as compared to reading from HDD, the benefit of prefetching is less obvious. One reason is the high I/O rate of SSD and relative low sampling rate of once per second by \emph{dstat}. However, benefits can still be seen by overlapping I/O and computation with prefetching.

In terms of batch size, the effect can be seen in Fig.~\ref{miniapp-batch}, where we vary the batch size and fix the number of threads to 8. Execution time for both with and without prefetch reduces as the batch size increases. This can be explained by higher utilization rate of GPU for parallelization.

\subsection{Checkpoint and Burst Buffer}

\begin{figure}[t]
	\begin{center}
		\includegraphics[width=0.5\columnwidth]{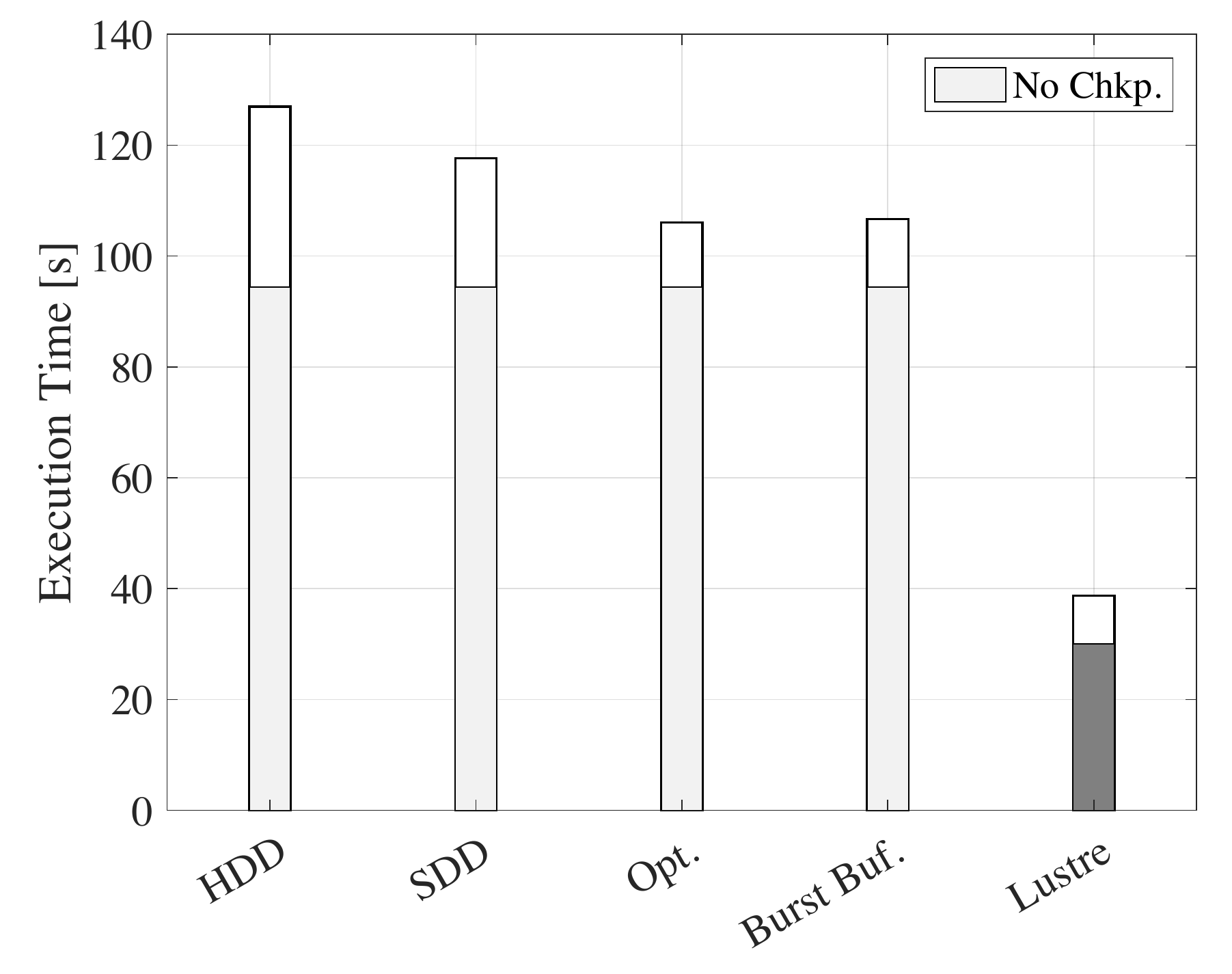}
		\caption{The median execution time of the extended mini-application in seconds, checkpointing to different storage devices. The gray area shows the same execution with no checkpoints being made. The tests on Tegner are shown in the dark gray bar.}
		\label{miniapp-checkpoint-runtime}%
	\end{center}
\end{figure}

We perform checkpoint performance test with our extended mini-application. 100 iterations are executed and one checkpoint is written and synchronized to disk every 20 iterations. Check-pointing of networks with large number of variables can be expensive, and since training must stop in favor for saving it can seriously hamper training speed. Top panel of Fig.~\ref{miniapp-checkpoint-runtime} shows the total run time of 100 iterations with five checkpoints to different storage media while the gray area represents the run time of 100 iterations without any checkpoint.

\begin{figure}[t]
	\begin{center}
		\includegraphics[width=0.6\columnwidth]{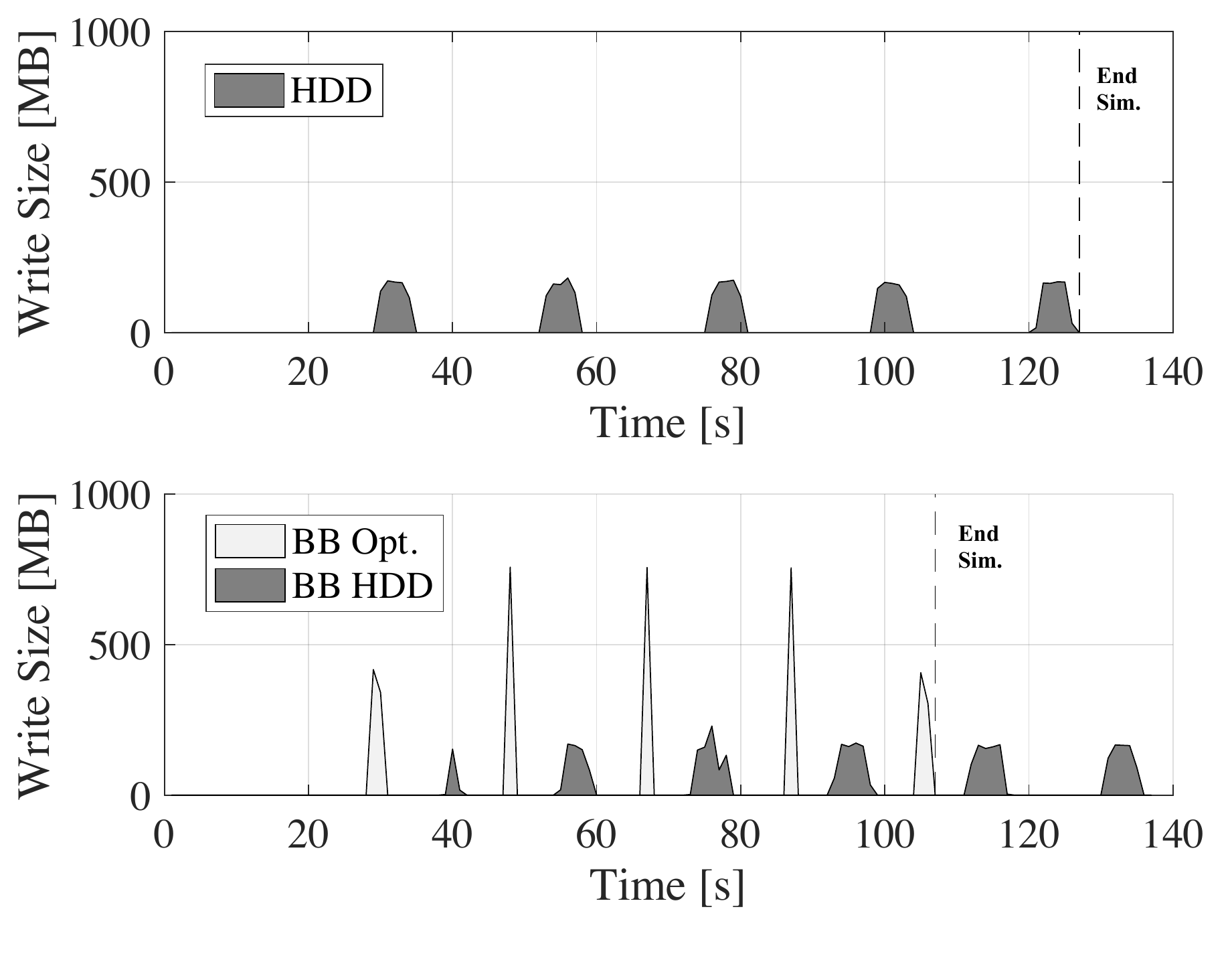}
		\caption{The top panel shows result from \emph{dstat} for checkpoint written to HDD. The bottom panel shows the case when checkpoint is written to Intel Optane as burst buffer and then moved to HDD. End of execution for the two cases are shown with dotted lines.}
		\label{miniapp-checkpoint-dstat}
	\end{center}
\end{figure}

Not surprisingly, checkpoint to Intel Optane gave the lowest run time on Blackdog, followed by checkpoint to SSD. Checkpointing to burst buffer also presents very favorable results, showing little difference then only checkpointing to Intel Optane without moving data to slower storage. The bottom panel of Fig.~\ref{miniapp-checkpoint-dstat} shows a time series of tracing collected by \emph{dstat}, which shows disk activity during execution of the mini-application. It can be seen that checkpointing to HDD uses considerable amount of time, thus hampering training performance. Checkpoints made to Intel Optane are shown in gray. It can be seen that Intel Optane performed extremely well in absorbing data written from the checkpoint process. Once the checkpoint is synchronized, a process is spawned to perform copying of the checkpoint files to HDD. The underlying file system of HDD ext4 performs caching of disk activities and flushes journal information and data to disk when necessary. Thus data written is not flushed to disk immediately, but when caches are flushed by the Operating System. Disk activity from copying to HDD is represented in gray areas, which are delayed comparing to the writes to Intel Optane and the flushing continues after the application ends. Since data is already written to Intel Optane, it is not necessary to enforce immediate synchronization of the checkpoint files when moved to HDD. The tests show considerable improvement when burst buffers are used. Since these devices used for burst buffer are typically smaller in size, by moving these files to HDD for archiving it is possible to cleanup the buffer for other data.

\section{Related Work}
Improving training time of DL networks is an active research area~\cite{distdl-preprint}. Attempts are made to benchmark several state-of-the-art GPU accelerated deep learning software tools such as Caffe~\cite{jia2014caffe}, CNTK~\cite{seide2016cntk}, Tensorflow~\cite{abadi2016tensorflow} and Torch where training performance does not scale very well on many-core CPUs. However through the use of GPUs together with a high-speed communications infrastructure they are able to provide performance improvements~\cite{pmlr-v28-coates13}\cite{7979887}. Performance of different frameworks often vary with different neural networks~\cite{7979887}. Bottlenecks of various DL frameworks have been studied through performance characterization of different convolution algorithms and their implementations~\cite{7975270}. Layer wise execution times and the batch processing times are also studied to identify limiting factors of each framework. There has been an effort to mitigate the inefficiencies of cuDNN in selecting a convolution algorithm based on the workspace size, so that faster algorithms are used with tight workspace constraints~\cite{ucudnn}. In terms of computing frameworks, Spark~\cite{Zaharia:2016:ASU:3013530.2934664} is a successful example of scaling different machine learning networks~\cite{HARNIE2017409}~\cite{moritz2015sparknet}.

Analysis and optimization of I/O is of vital importance for scalable machine learning which employ large datasets. Parallel I/O on HPC systems is an active research area in terms of parallelizing data access ~\cite{thakur1998case}. I/O optimizations range from application-side collective I/O to network and disk level request scheduling on the file system~\cite{6877251}. Benchmark tools such as IOR are developed for characterizing I/O performance in HPC systems ~\cite{shan2007using}. Assess pattern is also an important aspect for optimization. ~\cite{8291932} and~\cite{178404} looks into data assess pattern of distributed memory systems and proposes a modified framework S-Caffe. S-Caffe is a result of co-designing Caffe framework and MVAPICH2-GDR MPI runtime~\cite{awan2017s}. Theano-mpi is a Theano based distributed training framework for DNNs based on MPI which is used to drive inter-process communication~\cite{10.1007/978-3-319-58943-5_64}. DeepIO is an I/O framework for training deep neural networks featuring RDMA-assisted in-situ suffling, input pipelining and entropy-aware opportunistic ordering ~\cite{zhuentropy}. The framework is benchmarked against the TensorFlow dataset API, and a portable API for TensorFlow is developed to leverage DeepIO on different storage systems.
\section{Discussion and Conclusion}\label{discussion}
TensorFlow is one of the most used programming frameworks for implementing DL networks. During the training of the network, a large amount of data, consisting of relatively small files, is read and intermediate results are saved as checkpoints at a given time. For these reasons, the performance of the TensorFlow I/O might impact the overall performance of the application and it is important to understand how TensorFlow performs and optimize it. This is the goal of our work.

Using a micro-benchmark, similar in nature to the STREAM benchmark, we measured the read bandwidth of TensorFlow parallel I/O on different devices. We showed that increasing the number of threads, the I/O, the bandwidth in read increases. On Blackdog benchmarking environment, reading from HDD with two, four and eight threads resulted in a 1.65$\times$, 1.95$\times$ and 2.3$\times$ performance improvement when comparing with the I/O performance with one thread. On Tegner, we observed a 7.8$\times$ increase of bandwidth when using eight threads.

One of the major results of this paper is that the TensorFlow prefetcher is a key factor for the input pipeline performance in TensorFlow. The whole input data is divided into smaller datasets, called batches. While a batch is used to train the DL network on the GPU, the next batch is loaded and pre-processed on the CPU when using the TensorFlow prefetcher. In our mini-application, we found a complete overlap of per-batch computation and I/O pipelines leading to a minimal impact of I/O performance on the overall performance of the mini-application. We note that the overlap of computation and I/O depends on how long the per-batch computation lasts. The use of simpler DL networks, deeper input pipeline, faster accelerators and a decreased batch size leads to a decrease of the computation phase and a consequent overlap of computation and I/O. However, in practice we found the computation for one batch with a rather simple DL network, such as AlexNet, spans over 1-2 seconds period in most of the benchmark configurations, e.g. batch sizes and different GPUs, allowing for a complete overlap of computation and I/O.

While the input pipeline has a minimal impact because of overlap of computation and input pipeline, we found that checkpointing of intermediate weights of DL network depends on the I/O device in use and impacts the overall performance. The reason for this is the large amount of data to be written, roughly 600 MB in the case of AlexNet, and the fact that TensorFlow currently does not support overlap of checkpointing and computation. In this, the performance of checkpointing is limited by the latency and bandwidth of the device in use. For this reason, we implemented a burst-buffer solution for checkpointing to fast storage with small capacity (Intel Optane) and asynchronously copy the checkpoints to slower and high-capacity HDD. This resulted in obtaining the I/O performance of using the fastest I/O device with a performance improvement of 2.6$\times$ with respect to checkpointing directly to HDD.

As future work, we plan to focus on two main research lines related to this work. We first intend to investigate the performance of TensorFlow I/O using distributed systems and TensorFlow distributed datasets. In this work, we focused on single-node I/O performance, albeit using parallel I/O and one parallel file system in one test environment. As a second future task, we intend to investigate the TensorFlow I/O performance using object-store for HPC~\cite{chien2018exploring}, such as local installations of Ceph~\cite{weil2006ceph} and Seagate's Mero~\cite{narasimhamurthy2018sage}. This work requires the implementation of a TensorFlow file systems in the runtime, already supporting other remote object stores, such as AWS and Google Cloud.

Overall, we showed that the current TensorFlow provides high-performance I/O on single node in the data-ingestion phase and that burst-buffer can be an effective technique for fast check-pointing.

\section*{Acknowledgment}
Funding for the work is received from the European Commission H2020 program, Grant Agreement No. 801039 (EPiGRAM-HS) and Grant Agreement No. 800999 (SAGE2). Experiments were performed on resources provided by the Swedish National Infrastructure for Computing (SNIC) at PDC Center for High Performance Computing.

\bibliographystyle{IEEEtran}
\bibliography{tfIO}

\end{document}